*2024-2025 CRA Quadrennial Paper*

# The Post-Quantum Cryptography Transition: Making Progress, But Still a Long Road Ahead

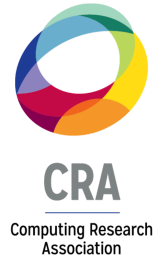
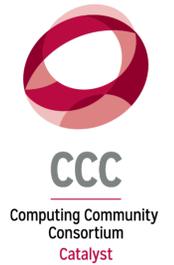

Brian LaMacchia (Farcaster Consulting Group), Matt Campagna (Amazon Web Services), William Gropp (University of Illinois Urbana-Champaign)

**The development of quantum computing threatens the security of our currently widely deployed cryptographic algorithms. While significant progress has been made in developing post-quantum cryptography (PQC) standards to protect against future quantum computing threats, the U.S. government's estimated $7.1 billion transition cost for non-National Security Systems alone, coupled with an aggressive 2035 deadline, will require sustained funding, research, and international coordination to successfully upgrade existing cryptographic systems.**

## Understanding the PQC Challenge

The confidentiality and integrity of our digital computer networks, communications, and systems depend critically on the science of cryptography. We rely on a small set of widely implemented and deployed cryptographic algorithms to secure communications among billions of Internet users daily, and it is public-key cryptography that enables two parties who have never communicated before to establish a secure and private communication channel over the Internet. Public-key cryptography also serves as the foundation for digital signatures, which allow us to digitally sign documents, ensure the integrity of software updates, and create digital forms of identity documents like e-passports.

Since 1994, when Peter Shor published his quantum factoring algorithm, we have known that the security of all our widely deployed public-key cryptographic algorithms would be threatened by the construction of large-scale, fault-tolerant quantum computers. Progress in quantum computing science led the NSA to announce in 2015 the need to begin transitioning our uses of public-key cryptography to newer public-key algorithms resistant to attacks even when aided by a quantum computer. These new algorithms are collectively known as "quantum-resistant cryptography" or, more commonly, "post-quantum cryptography" (PQC). NSA's call to begin the transition to PQC algorithms prompted the National Institute of Standards and Technology (NIST) in 2017 to launch a multi-year standardization effort to select new PQC cryptographic algorithms for standardization and eventual use throughout the U.S. government's IT systems.

CRA's [2020 Quadrennial Paper](#) on this topic reviewed the progress made since NIST launched its PQC standardization activity and urged organizations to begin immediate work on their PQC transition plans while awaiting NIST's final selections and the publication of the corresponding Federal Information Processing Standards (FIPS) specifications. Significant progress has been made on the road to PQC standardization since 2020, the most noteworthy milestones being:

1) In July 2022, NIST chose the first set of PQC algorithms for standardization, including one encryption algorithm and three digital signature algorithms. FIPS for three of these algorithms — [FIPS 203](#), [FIPS 204](#), and [FIPS 205](#) — were issued in July 2024, and the FIPS for the fourth algorithm is expected to be published within 18-24 months.

2) In September 2022, NIST issued a Request for Additional Digital Signature Schemes. NIST received 40 submissions, and in October 2024, 14 of these were selected to advance to Round 2 of the evaluation. This "second call" for signature algorithms is expected to yield one or more additional standardized algorithms, and the selection process is expected to take at least another two years.

3) Also in September 2022, NSA released the [Commercial National Security Algorithm Suite 2.0](#), incorporating a subset of NIST's selections into its requirements for National Security Systems (NSS) and related assets.

Additionally, both the executive and legislative branches have taken actions to further the PQC transition for the government and the private sector. In May 2022, President Biden issued [National Security Memorandum 10](#) (NSM-10), directing "specific actions for agencies to take as the United States begins the multi-year process of migrating vulnerable computer systems to quantum-resistant cryptography…with the goal of mitigating as much of the quantum risk as is feasible by 2035." NSM-10 required government agencies to produce inventories of their IT systems that are vulnerable to quantum computers within one year, with a focus on "high-value assets" and "high-impact systems." It also set a deadline of one year from the release of the first set of NIST PQC standards for agencies to develop a plan to upgrade their non-NSS IT systems to quantum-resistant cryptography.

Congress has also passed legislation related to the government's migration to PQC. In December 2022, Congress passed, and President Biden signed, the [Quantum Computing Cybersecurity Preparedness Act](#), directing the Office of Management and Budget (OMB) to:

1) Prioritize the acquisition and migration of federal agencies' information technology to post-quantum cryptography,

2) Create guidance for federal agencies to assess critical systems one year after the NIST issues planned post-quantum cryptography standards, and

3) Send an annual report to Congress that includes a strategy for addressing post-quantum cryptography risks, the funding necessary, and an analysis of whole-of-government coordination and migration to post-quantum cryptography standards and information technology.

This past July, OMB delivered its [first annual report on the PQC migration](#) to Congress. In its report, OMB estimated the funding necessary to upgrade the government's non-NSS IT systems to post-quantum cryptography at $7.1 billion.

## The Long Road Ahead

OMB's $7.1 billion top-line estimate highlights the scale of the collective challenge we face in upgrading all of our uses of public-key cryptography. By law, OMB's 2024 estimate covers only non-NSS federal government systems. When adding estimated costs for NSS systems, state and local government agencies, private sector systems, and civil society, the total cost will be significantly higher. Furthermore, the transition timetable mandated by NSA for NSS and by NSM-10 for other government systems is quite aggressive, particularly given the time it took both the public and private sectors to complete previous cryptographic algorithm transitions — which were simpler than the move to PQC. Thus, for the U.S. government to meet the deadlines set by NSA and NSM-10, Congress will need to fully and consistently fund each agency's PQC transition projects. These are significant budgetary line items, and OMB's initial cost estimates will likely change each year as the scope of the necessary work is refined. Nevertheless the work must be done to maintain the cryptographic security of our computing and communication systems.

At the same time, we must continue basic research on PQC cryptosystems and the underlying number-theoretic problems on which they are based. Cryptographic algorithms weaken and break over time, both because attackers have access to ever-increasing amounts of computing power and because new research can, at any moment, fundamentally change our perception of what is cryptographically secure. The need for research into an algorithm only increases once it is selected as a standard or even a finalist for future standardization. We already have a noteworthy example from the NIST PQC standardization process highlighting the importance of continued research: SIKE, a well-regarded PQC candidate encryption algorithm, was catastrophically broken by new fundamental research only weeks after being advanced by NIST to the fourth round of the competition. All the algorithms selected or to be selected by NIST for standardization are new and young constructions (even if they are based on old, well-known, number-theory problems), and we gain confidence in algorithms over time by trying to attack them. Thus, it is imperative that Congress and the Executive Branch continue to fund basic research into PQC even as the engineering work to transition government systems to PQC ramps up.

Additionally, for the PQC transition to be successful, NIST and other U.S. government stakeholders must continue to receive funding to coordinate with foreign governments and international standards-setting organizations. Most security protocols that must be updated to PQC algorithms are maintained by international standards organizations, such as the Internet Engineering Task Force (IETF) and the International Organization for Standardization (ISO). Upgrading these security protocols to PQC is a labor-intensive process, and early engagement is key to ensuring that national and international standards converge on common algorithms. Failure to invest in PQC standards coordination could lead to diverging international standards, which, in turn, would complicate the implementation of PQC-compliant security protocols by the private sector.

Finally, given the strong dependence of many U.S. government systems on private sector hardware and software infrastructure, government stakeholders must also ensure that the private sector transitions to PQC on time. While private sector organizations played a significant role in the initial NIST competition and in helping launch the Migration to Post-Quantum Cryptography project at the National Cybersecurity Center of Excellence, there is a risk that the current focus on investments in generative AI within the private sector may distract them from making timely and necessary investments in PQC for their own infrastructure. Government agencies should clearly communicate their specific PQC transition timelines to their technology providers, as well as when private sector vendors must complete the various phases of PQC transitions to meet government mandates and needs.

## Recommendations

In order to continue making steady progress over the next decade toward the goal of completing the PQC transition for U.S. government systems by 2035, we offer the following recommendations:

- Fully fund OMB's initial estimate for non-NSS IT transition costs, and require OMB to update that estimate annually as part of its PQC migration reports to Congress.

- Continue funding basic research on PQC algorithms and cryptosystems, with a particular focus on the hard problems underlying the new standards.

- Ensure that NIST and other U.S. government stakeholders have the necessary resources to continue PQC coordination with foreign governments and international standards-setting organizations.

- Consider expanding the scope of OMB's PQC migration reports to include relevant PQC transition timelines and information about necessary deliverables from the major technology providers for each government agency.

We have made significant progress over the past nine years on the transition to post-quantum cryptography, but the announcement of the first PQC algorithm FIPS marks just the conclusion of the initial phase of that journey. Maintaining the necessary pace will require consistent investments of time, expertise, and funding from all stakeholders.

---

*This quadrennial paper is part of a series compiled every four years by the **Computing Research Association (CRA)** and members of the computing research community to inform policymakers, community members, and the public about key research opportunities in areas of national priority. The selected topics reflect mutual interests across various subdisciplines within the computing research field. These papers explore potential research directions, challenges, and recommendations. The opinions expressed are those of the authors and CRA and do not represent the views of the organizations with which they are affiliated.*

*This material is based upon work supported by the U.S. National Science Foundation (NSF) under Grant No. 2300842. Any opinions, findings, and conclusions or recommendations expressed in this material are those of the authors and do not necessarily reflect the views of NSF.*